# Laser acceleration of highly energetic carbon ions using a double-layer target composed of slightly underdense plasma and ultrathin foil

W. J. Ma[1,3*], I Jong Kim[2, 4#], J. Q. Yu[1], Il Woo Choi[2, 4], P. K. Singh[2], Hwang Woon Lee[2], Jae Hee Sung[2, 4], Seong Ku Lee[2, 4], C. Lin[1], Q. Liao[1], J. G. Zhu[1], H. Y. Lu[1], B. Liu[5], H. Y. Wang[1], R. F. Xu[1,6], X. T. He,[1] J. E. Chen,[1] M. Zepf[6,7], J. Schreiber[3,5], X. Q. Yan[1, 8, †], and Chang Hee Nam[2, 9, ‡]

[1]*State Key Laboratory of Nuclear Physics and Technology, and Key Laboratory of HEDP of the Ministry of Education, CAPT, Peking University, Beijing, China, 100871*
[2]*Center for Relativistic Laser Science, Institute for Basic Science, Gwangju 61005, Korea*
[3]*Fakultät für Physik, Ludwig-Maximilians-Universität München, D-85748 Garching, Germany*
[4]*Advanced Photonics Research Institute, Gwangju Institute of Science and Technology (GIST), Gwangju 61005, Korea*
[5]*Max-Planck-Institute für Quantenoptik, D-85748 Garching, Germany*
[6]*Helmholtz-Institut-Jena, Fröbelstieg 3, 07743 Jena, Germany*
[7]*Department of Physics and Astronomy, Centre for Plasma Physics, Queens University, Belfast BT7 1NN, United Kingdom*
[8]*Collaborative Innovation Center of Extreme Optics, Shanxi University, Taiyuan, Shanxi, China, 030006*
[9]*Department of Physics and Photon Science, GIST, Gwangju 61005, Korea*
[#]*Present address: Optical Instrumentation Development Team, KBSI, Daejeon 34133, Korea*

**Abstract：**

We report the experimental generation of highly energetic carbon ions up to 48 MeV per nucleon by shooting double-layer targets composed of well-controlled slightly underdense plasma (SUP) and ultrathin foils with ultra-intense femtosecond laser pulses. Particle-in-cell simulations reveal that carbon ions residing in the ultrathin foils undergo radiation pressure acceleration and long-time sheath field acceleration in sequence due to the existence of the SUP in front of the foils. Such an acceleration scheme is especially suited for heavy ion acceleration with femtosecond laser pulses. The breakthrough of heavy ion energy up to multi-tens of MeV/u at high-repetition-rate would be able to trigger significant advances in nuclear physics, high energy density physics, and medical physics.

**Main text:**

Dense, energetic heavy ion bunches with ultrashort duration are highly demanded for high-energy-density physics and nuclear astrophysics [1,2]. Nearby the source, laser-driven ion acceleration can deliver exceptional ion bunches $10^{10}$ times denser than classically accelerated ion bunches [3,4], which highlights its application prospect in related fields. So far energetic heavy ions up to 80 MeV/u have been generated through Breakout Afterburner (BoA) [5,6] and Relativistic transparency (RT) [7,8] acceleration schemes. But both of the schemes require expensive 100s J level long-pulse lasers which were unable to operate at high repetition rate yet. Femtosecond laser pulses have been successfully applied in proton acceleration with the advantages of lower request on laser energy and Hertz-level repetition rate[9]. However, heavy ion acceleration with femtosecond pulses has not achieved the same success. The maximum energy per nucleon are still no more than 25 MeV/u, mostly only a few MeV/u [9-14], inefficient to overcome the coulomb barrier to excite nuclear reactions or isochronically heat bulk matters to warm dense state.

For femtosecond pulses, Target Normal Sheath Acceleration (TNSA) [15] and Radiation Pressure

Acceleration (RPA) [16,17] are the most widely employed schemes. In the TNSA scheme, the acceleration field (sheath field), established by laser-produced dilute thermal electrons, is easy to be diminished by contaminated protons and poorly ionized heavy ions which appear since the beginning of the interaction. Thus the acceleration of highly ionized heavy ions is strongly suppressed [18-21]. By completely removing the protons in the contamination layer, the energy of heavy ions can be improved to, in the best cases, a few MeV/u [18,19,21], which is still much lower than the maximum proton energy of 85 MeV achieved in TNSA scheme[22]. Compared to TNSA, RPA by using nanometer-thin foils as targets has been proven more beneficial to accelerating heavy ions due to the fact that the majority of bulk electrons in the targets are displaced by the radiation pressure. In particular, quasi-monoenergetic heavy ions can be obtained by entering light-sail RPA regime when circular polarized pulses and matching ultrathin foils are used. Experimental results show that carbon ions up to 25 MeV/u can be generated in the light-sail RPA regime [23]. The major problem for heavy ion acceleration in RPA scheme at current intensity is the fast decline of the acceleration field after laser reflection and unwanted early termination of acceleration due to plasma instability [24,25]. To generate highly energetic heavy ions, increasing the on-target laser intensity, or, prolonging the effective acceleration time are two ways to go.

Recently, a plasma-lens-enhanced RPA (PLE-RPA) scheme was realized [26,27] by putting a few-µm-thick and slightly overdense plasma (SOP) slab (plasma lens) right in front of an ultrathin foil. After propagating through the plasma lens, the laser pulses were strongly focused and steepened due to relativistic nonlinearity, creating ideal conditions for RPA. It turned out that the maximum energy of $C^{6+}$ can increase by a factor of 2.7 by adding a plasma lens, and quasi-monoenergetic $C^{6+}$ bunches were generated when circular polarized pulses were used. The shortcoming of PLE-RPA is that 30%-50% pulse energy was lost in SOP without significant contribution to the ion acceleration process. One can envisage that an efficient use of the lost energy would further boost the energy of ions.

In this work, we demonstrate the realization of a cascaded acceleration (CA) scheme especially suited for heavy ions. It happens when a laser pulse is focused on a target composed of homogeneous, tens-of-µm-thick, slightly underdense plasma (SUP) slab in front of an ultrathin foil. The targets have the same structure as in RLE-RPA scheme except the density of the near-critical-density layer is 10 times lower. Fig. 1(a) schematically illustrates the scheme. The density of the SUP is in the range of $0.1$-$1n_c$ to ensure that, direct laser acceleration, instead of wakefield acceleration, happens [28-30], where $n_c = m_e \omega^2 \varepsilon_0 / e^2$ is the critical density of plasma. In the SUP, the electrons are trapped in the plasma channel by electrostatic field and self-generated magnetic field, and in-phase accelerated by the laser field to energy far beyond the ponderomotive limit [31]. Because of their large γ-factor, the forward velocities of these "supperponderomotive electrons" are lower than the group velocity of the laser pulse, resulting in a dense and energetic electron flow behind the pulse. Once the laser pulse, self-steepened in SUP and followed by the electron flow, arrives at the ultrathin solid foil, ions residing in the foil undergo radiation pressure acceleration at first, then cascaded acceleration in a long-life-time sheath field dominated by the supperponderomotive electron flow. The superiority of CA is that RPA stage gives rise to an efficient ionization and pre-acceleration of highly ionized ions, and the TNSA stage thereafter ensures a sufficient long acceleration time. So the advantages of the two schemes are combined. Experimental results show that carbon ions with energy up to 48 MeV/u can be generated by using double-layer targets, which is, to our knowledge, about 2 times of the previous record obtained by using femtosecond lasers.

The experiments were performed using the PW Ti:sapphire laser facility at Center for Relativistic Laser Science (CoReLS) of Institute for Basic Science (IBS) in Korea. After a recollimating double plasma mirror (DPM) system, 33fs s-polarized laser pulses with energy of 9.2 J were focused to spots of 4.5 μm diameter (FWHM) using a f/3 off-axis parabolic mirror, resulting in a peak intensity of $5.5\times10^{20}$ W/cm$^2$, corresponding to a relativistic normalized vector potential of $a_0 = eE/m_e c\omega \approx 16$. After the DPM, the contrast of laser pulses was about $3\times10^{-11}$ at 6 ps before the main pulse, which is good enough to avoid the premature expansion of the nanotargets before the arrival of the main pulse. The incident angle was 2.4°, and the ion energy spectra were measured with a Thomson parabola (TP) placed in the direction of the laser axis. A microchannel plate (MCP) with a phosphor screen was equipped in the TP to convert the ion signal to optical signal imaged by a 16-bit CCD. The absolute response of the MCP was calibrated following the literature [32]. A 6-mm-thick tungsten plate with a 375-μm iris was used as the ion collimator in front of the TP, corresponding to the acceptance angle of $3.5\times10^{-8}$ steradians (sr). Bright and stable zero points, on top of halos resulted from secondary radiation from the collimator, were observed on CCD as shown in Fig. 1 (b). The carbon and proton traces converge at the zero point for every shot. The energy measurement error of the TP, estimated by considering the linewidth of the ion trace and the spatial resolutions of MCP and CCD, was about ±2.2 MeV for 60 MeV protons and ±3.3 MeV/u for 50 MeV/u C$^{6+}$.

The SUP layer of the double-layer targets was made of carbon nanotube foam (CNF) [33]. Such foam material can be synthesized through chemical vapor deposition (CVD) with bulk density in the range of 1.5-30 mg/cm$^3$ and sub-micrometer-scale homogeneity. Behind the CNF, a nanometer-thin solid foil made of diamond-like carbon (DLC) [34] was attached, in which only minute amount of protons lie in the contamination layer. In the experimental campaign, the bulk density of

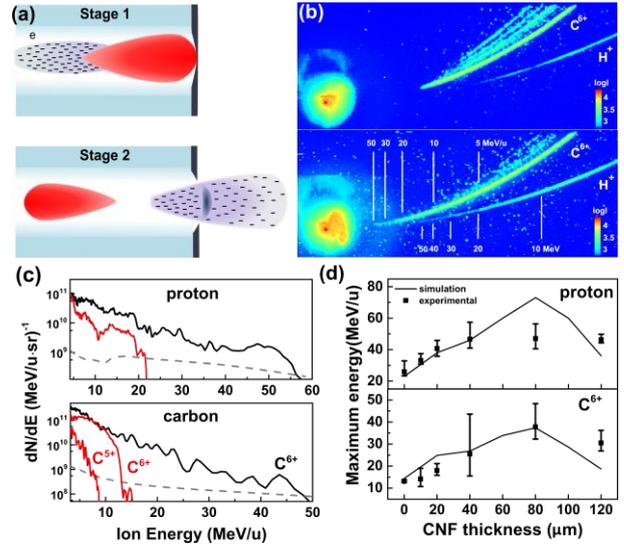

FIG. 1 (a) Schematic drawing of the cascaded acceleration process. (b), (c) Raw data and ion spectra obtained from a 20nm DLC target (upper image in (b), red lines in (c)) and a double-layer target with 80 μm CNF (lower image in (b), black lines in (c)). The dashed lines in (c) shows the detection threshold. (d) The dependence of the maximum proton/carbon energy on the thickness of CNF layer.

employed CNF was 3±1.5 mg/cm$^3$, corresponding to electron density of 0.4±0.2 $n_c$ if carbon atoms were fully ionized. The thickness of CNF was varied from 0 to 120 μm in the experiments, while DLC was fixed to 20 nm for all the targets. The raw data and ion spectra of 2 shots, obtained from a single-layer 20 nm DLC target and from a double-layer target with 80 μm CNF are shown in Fig. 1(b) and 1(c) respectively. Two features are obvious: (1) the energy and the number of carbon and proton ions obtained from the double-layer target are remarkably higher than those from the single-layer target. (2) 6+ is the *only* dominant charge state of carbons ions for the double-layer target, while multiple charge states were observed in the case of the single-layer target. For further confirmation, additional 53 shots by varying the thickness of CNF were made in the campaign. It turned out that the above features were repeatedly observed. The missing low charge states of carbon ions for the double-layer targets implies the ionization processes were not evolving but abrupt and complete, which is consistent with the fact that the pulses are self-steepened after propagating through the SUP. The maximum energies of protons and of $C^{6+}$ are plotted as a function of CNF thickness in Fig. 1(d), where the error bars reflect the shot-to-shot fluctuation and the dots are the arithmetic means. A strong dependency of ion energy on the CNF thickness is observed. The optimal thickness is 40/80 μm for proton/carbon acceleration, resulting in maximum 58/48 MeV/u, respectively. The solid lines in Fig. 1(d) depicts the numerical simulation results, which fit to the experimental results very well except for the case of proton from targets with 80 μm CNF. This discrepancy may be due to the fluctuation of the registered spectra considering the small number of high-energy ions entering the TP.

To illustrate the physics, 2D particle-in-cell simulations were performed using the EPOCH2D [35] code. The simulation window was $W_x \times W_z = 160\mu m \times 40\mu m$ with the cell size of dx=dz=10 nm. The laser pulse travelled along x from the left side with $\sin^2$ temporal profiles. Its peak laser intensity, focused spot and duration, is $5.5 \times 10^{20}$ W/cm$^2$, 4.5 μm and 33 fs, respectively. The electron density of CNF/DLC layer was $0.2n_c/50n_c$. The thickness of DLC layer in simulation was set to 200 nm to ensure its areal density is the same as that of 20 nm unionized DLC. The thickness of CNF varied from 0 μm to 120 μm. 10/1204 macro-electrons were placed into each cell of CNF/DLC.

Snapshots of $E_y$, $E_x$, and $(\gamma - 1)n_e$ at 3 different times obtained from simulations for targets with 60 μm CNF are shown in Fig. 2, where $E_y$ and $E_x$ are the electric fields, $\gamma$ and $n_e$ the $\gamma$-factor and the density of electrons from CNF, respectively. At T=430 fs, after propagating ~55 μm in CNF, the laser pulse duration (FWHM) is reduced from 33 fs to 15 fs with a steep rising edge due to relativistic nonlinearity in SUP. But in contrast to the case of PLE-RPA where the pulse is strongly self-focused in SOP, the intensity here is enhanced merely by 50% instead of by several times. A major portion of the laser energy is coupled to superponderomotive electrons in the NCD channel through direct laser acceleration [31,36], forming an electron flow with length of 15 μm behind the pulse. Once the steepened laser pulse reaches the solid foil (the blue dash dot line at 122 μm) at T=450 fs (Fig. 2(b)), RPA starts to dominate the acceleration process firstly. Electrons in the foil are piled up by the radiation pressure, resulting in a strong and localized charge separation field. A large number of carbon ions are abruptly ionized to the highest charge state and ripped off from the foil. At T=480 fs (Fig. 2(c)), the laser pulse has been reflected. The remaining acceleration is TNSA. The sheath field is established by the superponderomotive electron flow from the SUP and the thermal electrons from the DLC. The energy spectra of the two kind of electrons are shown in Fig. 2(d). It can be seen that both the number and the energy of the superponderomotive electrons are much higher than those of the thermal electrons. Thus the sheath acceleration stage is dominated by the superponderomotive electrons, which is remarkable different from the hybrid-RPA scheme where the sheath acceleration is purely due to thermal electrons [11,37]. Because the flow is dense and the electrons are highly energetic, the strong sheath field can last for a very long time. Fig. 2(e) shows that the major acceleration of carbon ions happens after the reflection of the laser pulse and last over 200 fs. Considering their low charge-to-mass ratio, such a stable long-time acceleration is crucial for achieving the efficient acceleration of heavy ions.

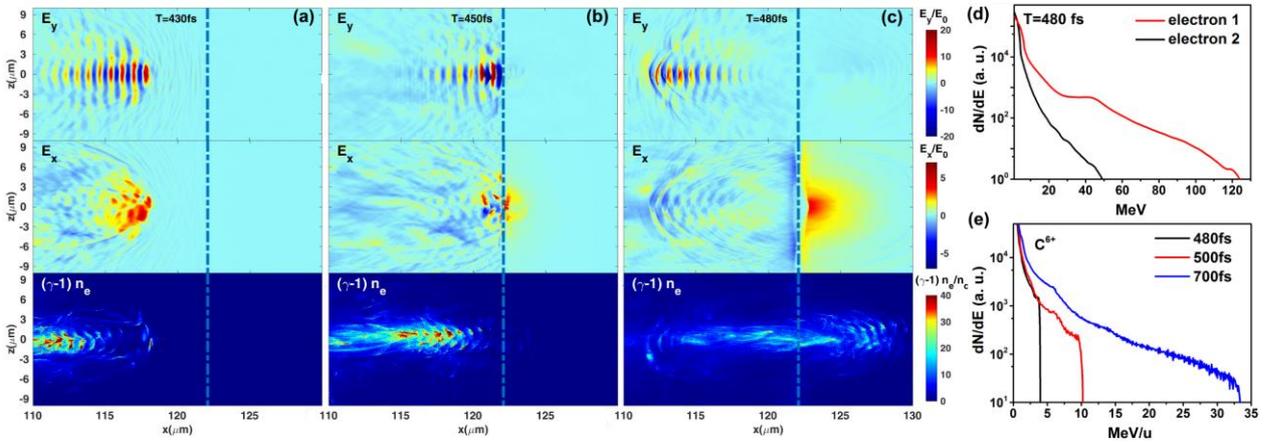

FIG. 2. Snapshots of transverse electric field ($E_y$), longitudinal electric field ($E_x$), and the energy density (($\gamma - 1)n_e$) of electrons from CNF at 430 fs (a), 450 fs (b), and 480 fs (c). (d) Energy spectra of electrons in CNF (red line) and in DLC (black line) at T=480 fs. (e) Ion spectra of carbons at different time.

As demonstrated in the experiments, there is an optimal thickness of the SUP layer for ion acceleration. Around this thickness, a significant amount of pulse energy is converted into electron flow and eventually contributes to the sheath field acceleration, meanwhile the remaining pulse is strong enough to displace the bulk electrons in the target. This is confirmed by simulations in Fig. 3(a) by tracking the energy gain rate (EGR) of the most energetic carbon ions. In the case of a single foil target where RPA dominates, the EGR starts to rise after the unshaped laser pulse arrives at 420 fs, then peaks at about 480 fs when the pulse is reflected, and quickly declines afterwards. In contrast, for the double-layer target with 60 μm CNF, the EGR, peaking at 500 fs when the electron flow passes through the foil, is much higher in magnitude and lasts much longer. If the CNF layer is too thick, for example 120 μm, the laser pulse is seriously depleted and filamented [38] before it reaches the DLC, so that RPA can not be triggered. The ions are then accelerated merely by the sheath field with a reduced strength. For a comparison to PLE-PRA regime, a simulation by setting the electron density of CNF to $2n_c$ and thickness to 6 μm was performed as well and presented in Fig. 3(a). The 10-fold increment of the CNF density to slightly overdense results in stronger self-focusing but without a long superponderomotive electron flow. As a result, although the EGR in RPA stage (before 480 fs) is higher, the final ion energy is lower compared to $0.2n_c$.

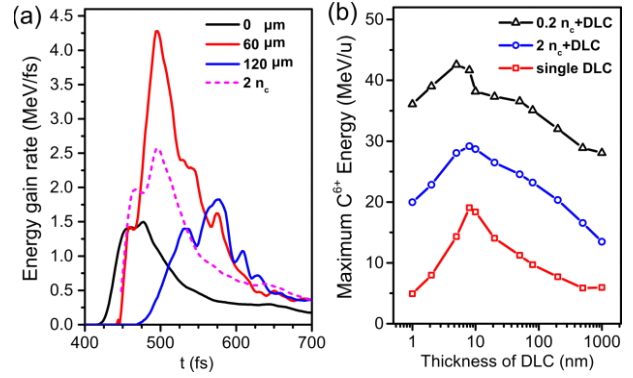

FIG. 3. (a) Energy gain rate (MeV/fs) of $C^{6+}$ ions as a function of time for targets with different CNF layers obtained from simulations. (b) The dependance of maximum $C^{6+}$ energy on the thickness of DLC for single DLC targets and double-layer targets.

Besides of the length and density of the SUP layer, simulations reveal that the thickness of the DLC foil imposes significant influence on the ion acceleration as well. Fig. 3(b) shows the dependence of the maximum $C^{6+}$ energy on the thickness of the DLC foils. In the case of $0.2n_c$ SUP, the optimal DLC thickness is 5 nm, and the maximum carbon energy vary little for 10 nm-100 nm DLCs and eventually drops to 27 MeV/u for 1 μm DLC. Such a dependency is different from the case of $2n_c$ where the optimal thickness is 10 nm and the maximum $C^{6+}$ energy quickly drops to 12 MeV/u. The simulation results clearly demonstrate the importance of using an ultrathin foils behind the foam, so that the RPA stage can efficiently ionize and pre-accelerate heavy ions. This has been proved by previous studies where foam-coated micrometer-thick metal foils were shot at laser intensity close to ours. The enhancement of proton energy by coating the foam was prominent there, but the energy of heavy ions were no more than 11 MeV/u [39,40]. As a comparison, the results from single layer DLC foils irradiated by linearly polarized laser pulses are also shown in Fig. 3(b). One can see that the maximum $C^{6+}$ energies, relying highly on the thickness of DLC, are significantly lower than those using the double-layer targets for all the cases.

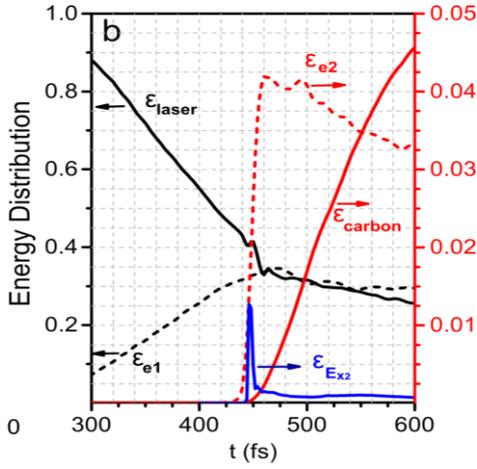

FIG 4. Normalized energy distribution among the laser pulse ($\varepsilon_{\text{laser}}$), electrons in CNF ($\varepsilon_{e1}$), electrons in DLC foil ($\varepsilon_{e2}$), longitudinal electrostatic field in the DLC foil ($\varepsilon_{E_{x2}}$), and carbon ions ($\varepsilon_{\text{carbon}}$) as a function of time.

For the case of the double-layer target with 60 μm $0.2n_c$ CNF, FIG. 4 depicts energy distribution, normalized by the initial laser energy, among the laser pulse ($\varepsilon_{\text{laser}}$), electrons in CNF ($\varepsilon_{e1}$), electrons in DLC foil ($\varepsilon_{e2}$), carbon ions ($\varepsilon_{\text{carbon}}$), and longitudinal electrostatic field in the DLC foil ($\varepsilon_{E_{x2}}$) with respect to time, here $\varepsilon_{\text{laser}}$ is calculated by numerical integrating the transverse electromagnetic field of $\iint(\varepsilon_0 E_y^2 + B_z^2/\mu_0)dxdz$, and $\varepsilon_{Ex2} = \iint_{x=122\mu m}^{x=122.2\mu m} \varepsilon_0 E_x^2 dxdz$. It can be seen that the laser energy is gradually transferred to the electrons in the CNF with the propagation of the pulse in SUP. About 35% of laser energy is coupled into $\varepsilon_{e1}$ prior to the second-stage acceleration. After the laser pulse reaches the foil at T=450 fs, 4.2%/1.2% of the laser energy is converted into $\varepsilon_{e2}/\varepsilon_{E_{x2}}$ within 25 fs in the RPA stage, then partially transferred to ions. At T=600 fs when the acceleration is close to the end, $\varepsilon_{e1}$, $\varepsilon_{e2}$ and $\varepsilon_{E_{x2}}$ drops to 30%, 3.2% and 0.1% respectively, and $\varepsilon_{\text{carbon}}$ increases to 4.5%. Based on the changes of the energy distribution, it is reasonable to deduce that 70% of energy on carbon ions comes from the electron flow generated in CNF. In other words, the sheath acceleration plays a more important role than RPA from the energy conversion point of view. But the RPA stage is still important because the heavy ions can gain this large fraction only after they were efficiently ionized and pre-accelerated in RPA stage.

To reveal the dependency of ion energy on laser intensity, simulations were performed by varying laser intensity. The maximum carbon energies obtained from the simulations are shown in Fig. 5 as the solid and the dashed lines, where the laser energy is calculated as $\varepsilon_{\text{laser}}(J) = 1.69 \times I_0(W/cm^2)/10^{20}$ according to the relationship between intensity and laser energy in our experiments. The solid line is obtained at the optimal thickness of CNF for different intensity with a fixed CNF density of 0.2 $n_c$. The dashed line is obtained by scaling up the density of CNF with laser intensity as well. It can be seen that the carbon energy is higher in the latter case, following $E_{max} \propto I^{0.6}$. This scaling is superior to TNSA but inferior to RPA. By comparing with existing RPA and TNSA results, one can speculate that for laser intensity available nowadays and in near future, the cascaded acceleration would be a realistic optimal scheme for the generation of highly energetic heavy ions. It should be noted that the scaling obtained from carbon ions may not be directly applied to very heavy ions like Cu, Au, since the detailed ionization dynamics is not taken into account here. But the advantages of cascaded acceleration will be sustained.

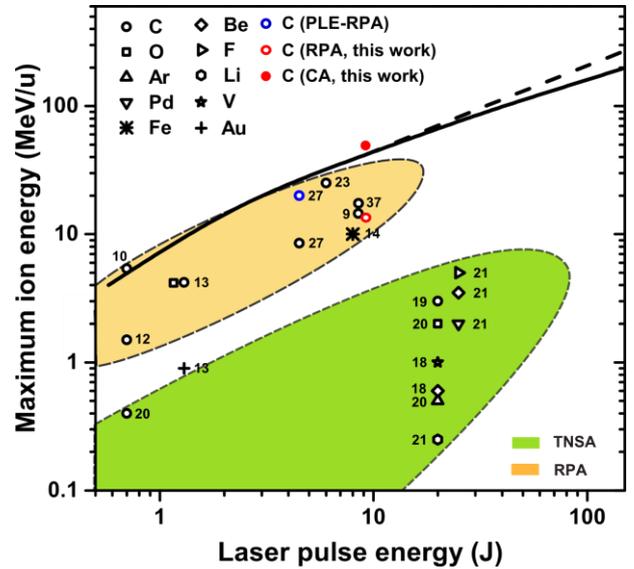

FIG.5. Summary of reported experimental results (shown by the reference number) and scaling of carbon ions in cascaded acceleration scheme.

In summary, we demonstrate that the cascaded laser acceleration of carbon ions can be achieved by combining a tens-of-micrometer-thick, slightly underdense plasma layer with a nanometer-thin foil. We identified a parameter range in which the subsequent interplay of RPA and sheath acceleration leads to substantially higher maximum ion energy. The scheme is especially suited for heavy ion acceleration. Experimental and simulation results confirm its superiority over other mechanisms for carbon acceleration at realistic laser parameters currently accessible.

Acknowledgements: The work has been supported by the Institute for Basic Science of Korea under IBS-R012-D1, the National Basic Research Program of China (Grant No.2013CBA01502, 11475010, 11775010, 61631001), NSFC (Grant Nos.11535001) and National Grand Instrument Project (2012YQ030142). J. Q. Yu wants to thank the Projects 2016M600007, 2017T100009) funded by China Postdoctoral Science Foundation. The PIC code Epoch was in part funded by the UK EPSRC grants EP/G054950/1. Our simulations were carried out in Max Planck Computing and Data Facility and Shanghai Super Computation Center. J. Schreiber and W.J.M. acknowledge support by the DFG-funded MAP cluster and the LMU target factory. W. J. Ma and I. J. Kim made equal contribution to this work in experiments.

Corresponding to:
*wenjun.ma@pku.edu.cn
†x.yan@pku.edu.cn
‡chnam@gist.ac.kr